\documentclass[conference]{IEEEtran}
\IEEEoverridecommandlockouts
\usepackage{cite}
\usepackage{caption}
\usepackage{subcaption}
\usepackage[implicit=false]{hyperref}
\usepackage{amsmath,amssymb,amsfonts}
\usepackage{algorithmic}
\usepackage{graphicx}
\usepackage{todonotes}
\usepackage{textcomp}
\usepackage{xcolor}
\usepackage{array}
\usepackage{url}
\usepackage{booktabs}
\usepackage{tabularx}
\usepackage{comment}

\usepackage{enumitem}
\setlist{noitemsep, topsep=2pt, parsep=2pt, partopsep=0pt}

\def\BibTeX{{\rm B\kern-.05em{\sc i\kern-.025em b}\kern-.08em
    T\kern-.1667em\lower.7ex\hbox{E}\kern-.125emX}}

\begin{document}
    
\bstctlcite{BSTcontrol}

\makeatletter
\newcommand{\linebreakand}{%
  \end{@IEEEauthorhalign}
  \hfill\mbox{}\par
  \mbox{}\hfill\begin{@IEEEauthorhalign}
}
\makeatother

\title{	MusiSPARQ: Subscription-Aware Adaptive Resource Control for Musical Metaverse Services\\\thanks{This work is supported by MUSMET project funded by the
EIC Pathfinder Open scheme of the European Commission
(grant agreement n. 101184379), and has received funding
from the Swiss State Secretariat for Education, Research and
Innovation (SERI). Corresponding author: ali.alhousseini@supsi.ch}
}

\author{  
\IEEEauthorblockN{
    Ali Al Housseini\IEEEauthorrefmark{1}\IEEEauthorrefmark{4},
    Omran Ayoub\IEEEauthorrefmark{1}, 
    Jaime Llorca\IEEEauthorrefmark{2}\IEEEauthorrefmark{3}\\
    }

  \IEEEauthorblockA{\IEEEauthorrefmark{1}
   University of Applied Sciences and Arts of Southern Switzerland, Lugano, Switzerland \\ \IEEEauthorrefmark{2}
    Centre Tecnologic de Telecomunicacions de Catalunya (CTTC/CERCA), Castelldefels, Spain
    \\ \IEEEauthorrefmark{3}
    New York University, New York, USA; \IEEEauthorrefmark{4}Università della Svizzera Italiana, Lugano, Switzerland}

}

\maketitle

\begin{abstract}

The Musical Metaverse relies on distributed audio, visual, inference, and delivery functions deployed across heterogeneous edge-cloud resources. Queue-aware orchestration frameworks such as SPARQ jointly optimize service placement, routing, and resource allocation, but fix the execution regime of each resource in advance: it is either shared, favoring statistical multiplexing, or guaranteed, favoring performance isolation. This paper introduces MusiSPARQ, a subscription-aware control framework that promotes resource isolation from a fixed infrastructure attribute to a runtime decision. Each switchable resource is represented through coupled shared- and guaranteed-resource pools drawing from the same physical-capacity budget, and at each orchestration epoch the controller selects which class-specific service flows remain shared and which receive isolated capacity. The per-epoch decision jointly accounts for class-specific latency objectives, per-class maximum-latency limits, queueing stability, isolation cost, and switching overhead. In a dynamic virtual-concert scenario, MusiSPARQ reduces the Premium latency-target violation rate from 16.19\% to 0.29\% relative to fully shared execution, while reserving on average only 0.71\% of capacity and, unlike static isolation, keeping Standard and Basic traffic within their latency targets.

\end{abstract}


\section{Introduction}
The Musical Metaverse (MM) enables geographically distributed performers and audiences to participate in shared and immersive musical experiences through interconnected audio, visual, and interactive services \cite{turchet2023musical, turchet2018internet}. Delivering interactive real-time experiences requires far more than transporting media streams between endpoints; it imposes stringent requirements on end-to-end latency, jitter, and packet loss, while demanding tight synchronization among participants.

A typical service pipeline may include multimodal data capture, AI-assisted audio enhancement, spatial-audio rendering, avatar-motion inference, scene generation, media encoding, and personalized audience delivery. These functions form interdependent processing chains that must be orchestrated across heterogeneous edge and cloud resources while preserving application-specific latency and synchronization requirements \cite{turchet20245g,mauro2026idago}.

Orchestrating these processing chains involves an inherent trade-off between resource sharing and performance isolation. Co-located service functions may execute on the same edge GPU, so that capacity left unused by one function can be exploited by another. This statistical multiplexing improves infrastructure utilization and reduces the capacity that must be reserved in advance, but it also couples performance: the delay experienced by one workload depends on the traffic and processing demands of the others sharing the resource. During a live MM event, for instance, a surge in avatar-rendering requests may increase the queueing delay of an audio-processing function deployed on the same edge server, even when the audio workload itself is unchanged. Isolated execution removes this interference by dedicating resource quotas to selected workloads, but reserved capacity may sit idle when demand is low and cannot be exploited by competing services. An effective orchestration mechanism must therefore decide not only where service functions are placed and how their traffic is routed, but also when the benefits of isolation justify the associated loss of sharing efficiency.

Recent works have investigated delay-aware placement, routing, and resource allocation for distributed edge applications \cite{poularakis2020mec,mauro2026idago}. Among them, SPARQ provides a queue-aware framework for jointly optimizing service-function placement, inter-function traffic routing, and communication-computation resource allocation for AI-intensive applications \cite{sparq}. SPARQ distinguishes between Guaranteed Resources (GR), where individual service commodities receive dedicated allocations and are modeled through independent M/M/1 queues (single-server queues with Poisson arrivals and exponentially distributed service times), and Shared Resources (SR), where heterogeneous commodities contend for a common resource pool and their coupled delays are represented through an M/G/1 queue (a single-server queue with Poisson arrivals and a general service-time distribution). This distinction is particularly relevant to the MM, whose service chains combine functions such as audio enhancement, motion inference, scene rendering, and media encoding. These functions may execute on common edge accelerators while exhibiting different processing demands and latency sensitivities. SPARQ therefore provides a suitable foundation for representing the nonlinear contention effects that arise in such deployments.

\begin{figure}[t]
    \centering
    \includegraphics[width=\linewidth]{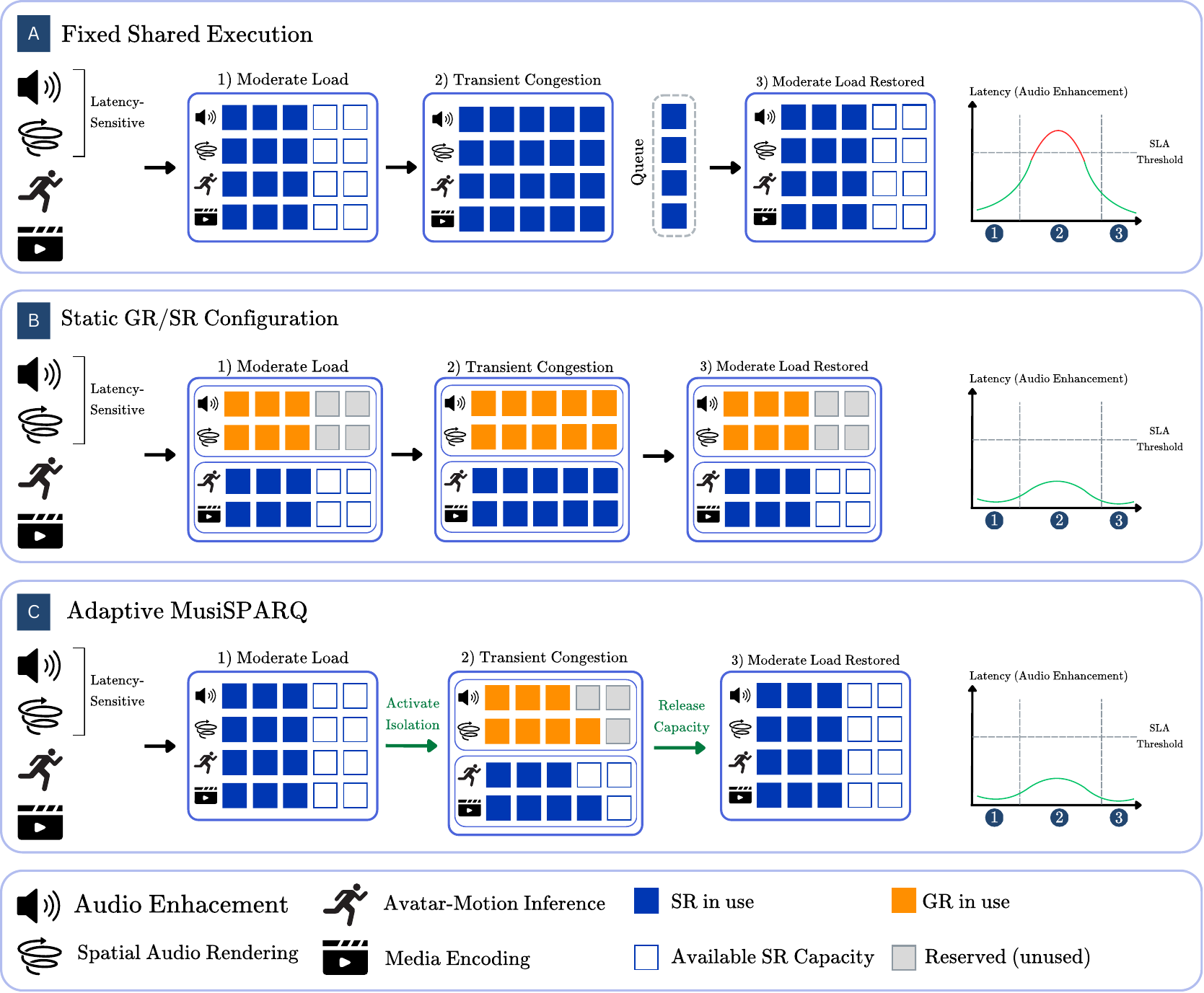}
    \caption{\footnotesize  Motivation for adaptive resource isolation. Fixed shared execution preserves resource-sharing efficiency but may violate the latency objectives of time-sensitive flows during transient congestion.
    A static GR/SR configuration protects these flows but leaves reserved capacity unused under moderate load. MusiSPARQ activates isolation
    selectively during congestion and returns the capacity to shared
    execution when the load subsides.}
    \label{fig:adaptive-isolation}
\end{figure}

In SPARQ, however, the GR or SR execution regime associated with an infrastructure resource is fixed during orchestration. The optimizer can determine service placements, traffic routes, and capacity allocations, but it cannot adapt the isolation regime as workload conditions evolve. This assumption becomes restrictive in subscription-based MM services, where service tiers may be associated with different latency objectives and where the number and composition of active users vary throughout a live event. Under moderate network load, shared execution may satisfy all latency requirements, making dedicated isolation unnecessary. During an audience surge, however, the same shared configuration may no longer protect latency-sensitive traffic. Conversely, permanently reserving isolated capacity for high-tier services may leave that capacity underutilized during low-demand periods, when it could be safely shared without violating their latency objectives.

Figure~\ref{fig:adaptive-isolation} distills this limitation into three operating regimes. (A) Fixed shared execution preserves statistical multiplexing under moderate network load, but transient congestion can push the latency of time-sensitive processing beyond its target. (B) A static GR/SR configuration instead provides stronger protection by reserving dedicated capacity for selected flows, but part of this reservation remains unused when demand subsides. The
desired behavior is therefore adaptive (C): shared execution should be retained while class-specific latency objectives are satisfied, isolation should be activated only for flows requiring stronger protection during congestion, and the reserved capacity should be returned to the shared pool once the demand surge has passed.

Motivated by this need, we introduce \emph{MusiSPARQ}, a subscription-aware control framework for MM applications that promotes resource isolation from a fixed infrastructure attribute to a runtime orchestration decision. MusiSPARQ is a controller that operates on top of a given service deployment, representing each switchable physical resource through logical GR and SR pools that draw from a common physical-capacity budget. At each orchestration epoch, the controller determines which component--class flows should remain in shared execution and which should
receive isolated capacity, according to the observed workload, class-specific latency objectives, and subscription requirements. Isolation is therefore selective and reversible: it is activated when contention threatens a protected flow and released when shared execution again becomes sufficient.

Figure~\ref{fig:system-overview} situates this decision process within the MM service environment. Performers and AI-based services are supported by a shared edge--cloud infrastructure serving users with differentiated service requirements. The MusiSPARQ controller observes workload
conditions, latency objectives, and subscription classes, and dynamically
selects between shared execution and stronger performance isolation.

\begin{figure}[t]
    \centering
    \includegraphics[width=0.8\linewidth]{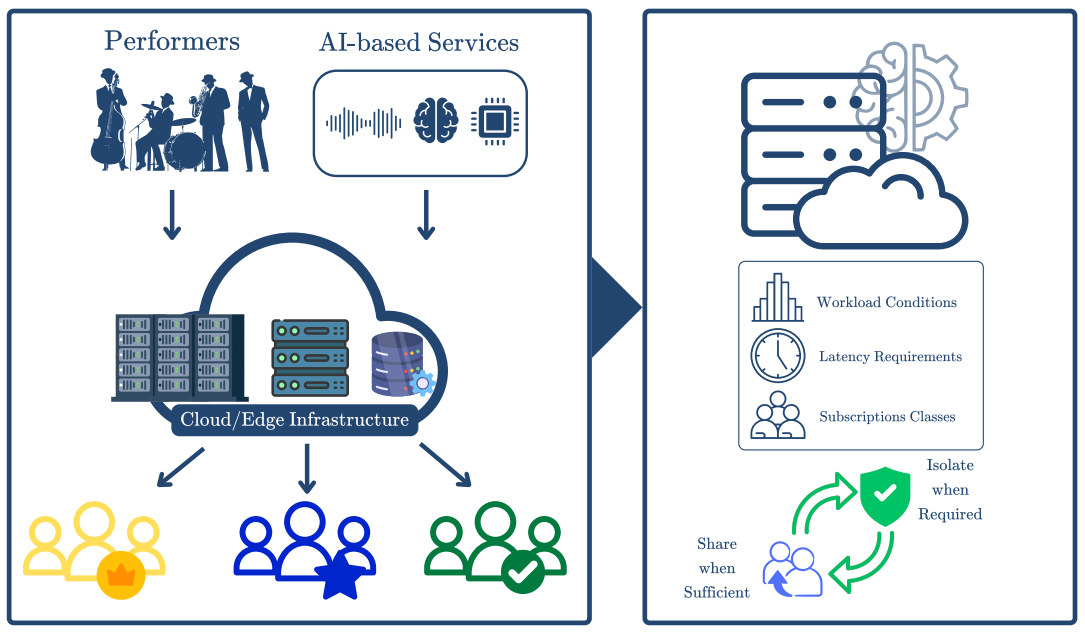}
    
    \caption{\footnotesize System-level overview of MusiSPARQ. The orchestration controller
    observes workload conditions, class-specific latency objectives, and
    subscription requirements, and determines whether service flows can remain
    in shared execution or require isolated capacity.}
    \label{fig:system-overview}
\end{figure}


The main contributions of this work are summarized as follows:

\begin{itemize}

\item \textbf{Adaptive resource-isolation model:}
We extend SPARQ by promoting the GR/SR execution regime from a fixed infrastructure attribute to a runtime decision. Class-specific service flows are assigned to logical GR or SR pools coupled through the fixed capacity of the underlying physical resource.

\item \textbf{Subscription-aware isolation formulation:}
We formulate the per-epoch isolation decision as an optimization problem that jointly accounts for class-specific latency objectives and violation weights, isolation cost, shared physical-capacity and stability constraints, and the overhead of switching between GR and SR execution. Per-class maximum-latency limits prevent the protection of one class from causing unbounded degradation of the others.

\item \textbf{Online orchestration and evaluation:}
We develop an exact per-epoch selection procedure and evaluate it under dynamic MM workloads against fully shared, static, and aggressive isolation baselines. 

\end{itemize}

The remainder of this paper is organized as follows. Section~\ref{sec:related_work} reviews related work. Section~\ref{sec:sparq_preliminaries} summarizes the SPARQ service representation and execution regimes on which MusiSPARQ builds. Section~\ref{sec:musisparq_system_model} formalizes the subscription-aware adaptive-isolation model and the per-epoch decision problem. Section~\ref{sec:experimental_setup} describes the experimental setup, Section~\ref{sec:results} discusses the results, and Section~\ref{sec:conclusion} concludes the paper.

\section{Related Work}
\label{sec:related_work}

The Internet of Musical Things (IoMusT) and Internet of Sounds (IoS) visions describe ecosystems where musical devices, stakeholders, services, and immersive environments exchange synchronized audio, visual, and control data \cite{turchet2018internet,turchet2023potential}, implemented in the MM through multi-stage chains such as audio enhancement, spatial rendering, motion inference, encoding, and personalized delivery \cite{turchet2023musical}. Latency, jitter, and stream misalignment are known to affect tempo stability, coordination, and presence, with acceptable latency depending on the musical task rather than a single universal threshold \cite{rottondi2016overview}. IoS service orchestration must therefore protect the end-to-end processing path while accounting for the differing latency requirements of its functions and users.

One way to address this challenge is by improving the communication path itself and moving delay-sensitive processing closer to users \cite{turchet2023latency,turchet20245g,turchet2024slicing}. Experiments with private fifth-generation (5G) networks show that modern systems can approach the latency demands of networked musical applications, though latency grows with the number of connected devices and background traffic, and short delay excursions remain possible even when the average latency stays within bounds \cite{turchet2023latency, turchet2023private5g, durre2022public5g}. Complementary architectures combine 5G with Software-Defined Networking (SDN) and Multi-access Edge Computing (MEC) to place or migrate immersive-audio functions dynamically according to traffic and network conditions \cite{centenaro2020iomust5g,turchet20245g,rinaldi2021spatialaudio,martusciello2023edgeaudio,centofanti2024taminglatency}, while network slicing allocates dedicated communication resources to musical traffic that requires stronger protection \cite{turchet2024slicing,turchet2024isccslicing}. Together, these studies establish that programmable networks and edge deployment can reduce transport delay and improve reliability. However, they also reveal the next bottleneck: once multiple audio, rendering, inference, and media functions are co-located on the same edge node, network-level prioritization alone cannot prevent them from interfering as they compete for the same processor or accelerator resources.

This remaining problem connects IoS research to edge--cloud service orchestration. Cloud Network Flow and stream-processing models jointly optimize function placement, routing, and communication--computation allocation \cite{barcelo2016iotcloud,michael2019stream}; MEC extensions add storage-aware service placement and request routing \cite{poularakis2020mec}, and IDAGO adapts this line to information-aware media graphs with branching, merging, scaling, and stream replication \cite{mauro2026idago}. Dynamic-control and VNF-orchestration work further adapts processing, routing, and resource decisions to time-varying demand \cite{feng2018dcnc,agarwal2019vnf}, while QoE- and subscription-aware systems translate user class into admission, priority, or minimum-rate policies \cite{seppanen2014qoe,shajaiah2014subscription}. These studies show how to place service pipelines and differentiate users, but they do not decide when a delay-sensitive musical function should remain in a shared queue and when it should receive isolated processing to avoid the sharp queueing-delay growth caused by saturation.

SLA-aware orchestration and network slicing explore the same sharing-versus-guarantees boundary at the tenant or slice level. Zheng \emph{et al.} combine guaranteed slice shares with competition for excess capacity to improve utilization without discarding service guarantees \cite{zheng2023greet}; Jo\v{s}ilo and D\'an study dynamic slice selection that jointly adapts wireless and edge-computing resources for latency-sensitive tasks \cite{josilo2022slice}; and SLA-ORECS forms isolated SLA channels from dedicated and shared edge--cloud resources and reconfigures them according to demand \cite{lan2024slaorecs}. These systems show that sharing and guarantees can coexist, but their abstractions remain tenants, slices, SLA channels, or independent tasks. They do not model how subscription-class flows traverse the same interdependent musical service graph, how their delays become coupled at shared processing stages, or how selectively isolating one component-class flow changes the residual capacity and queueing delay experienced by the others.

The closest foundation for resolving this limitation is SPARQ, which brings contention-aware delay modeling into service-graph orchestration \cite{sparq}. SPARQ captures the cost--delay trade-off between shared execution and guaranteed execution, thereby avoiding the overly conservative behavior of private-delay overprovisioning. Its limitation for the present setting is that the execution regime is fixed before optimization, i.e., resources are already classified as shared or guaranteed. Thus, SPARQ can optimize placement, routing, and provisioning under a prescribed sharing regime, but it does not decide when a particular subscription-class flow should move between shared and isolated execution as load and priorities evolve.

These strands point to a common requirement that remains open for MM services: the orchestration model must preserve the structure of an interdependent service graph, account for contention-induced queueing delay, and still make class-specific protection decisions at runtime. MusiSPARQ addresses this requirement by deciding at each epoch which component--class flows can safely remain shared and which should be isolated to protect subscription-specific latency objectives.

\begin{figure}
    \centering
    \includegraphics[width=0.5\linewidth]{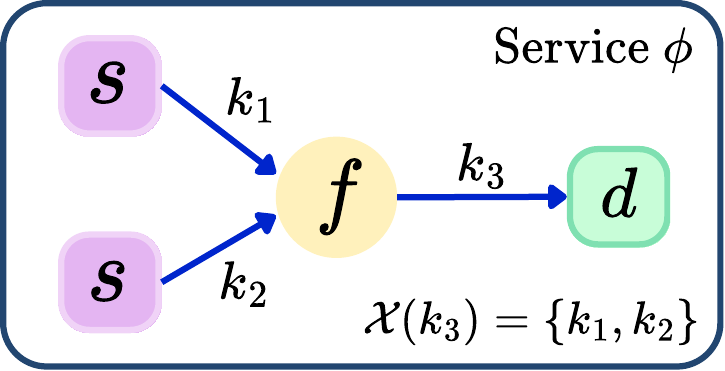}
    \caption{\footnotesize  Example of a service graph. Commodity $k_3$ is produced requiring as
input commodities $k_1$, $k_2$.}
    \label{fig:fig_exmpl}
\end{figure}

\section{SPARQ Preliminaries}
\label{sec:sparq_preliminaries}

\subsection{Service and Infrastructure Representation}

The physical edge--cloud infrastructure is represented by a directed
graph
$  \mathcal G=(\mathcal V,\mathcal E)$,
where $\mathcal V$ contains the network and compute-capable nodes,
and $\mathcal E$ contains the physical communication links.

A service $\phi$ is represented by a directed acyclic graph whose
vertices are service functions and whose directed edges are
\emph{commodities}. A commodity is the data stream produced by one
function and consumed by the next. We denote the commodity set by
$\mathcal K$. If commodity $j$ is an input to the function producing
commodity $k$, we write $ j\in\mathcal X(k)$. For example, in Fig.~\ref{fig:fig_exmpl}, commodities $k_1$ and
$k_2$ are processed by function $f$ to produce commodity $k_3$;
hence, $\mathcal X(k_3)=\{k_1,k_2\}$. In a MM
service, such commodities may carry captured audio, enhanced audio,
motion information, rendered media, or encoded content between
successive processing functions.

To represent function placement and routing within a common
model, CNF augments the physical graph with virtual production,
consumption, and computation nodes. The resulting cloud-augmented
graph is
$    \mathcal G^a=(\mathcal V^a,\mathcal E^a).
$
An augmented edge $(u,v)\in\mathcal E^a$ may therefore represent a
physical communication link, the execution of a service function on
a computation system, or the production or consumption of data.

For each commodity $k$ and augmented edge $(u,v)$, CNF defines
$f_{uv}^{k}\in\{0,1\}$ where $f_{uv}^{k}=1$ indicates that commodity $k$ uses edge $(u,v)$.
On a communication edge, this variable identifies part of the route
followed by the commodity. On a computation edge, it identifies the
system on which the function producing the commodity is executed.

SPARQ jointly optimizes these variables subject to flow-conservation,
service-dependency, and endpoint constraints, thereby determining
the placement of the service functions and the routes connecting
them~\cite{sparq}. MusiSPARQ operates after this optimization and treats the resulting solution as a fixed embedding:
\begin{equation}
    \bar{\mathbf f}
    =
    \{\bar f_{uv}^{k}:
      k\in\mathcal K,\,(u,v)\in\mathcal E^a\}
\end{equation}

\subsection{Resource-Sharing Regimes}

An augmented edge may expose several resource types
$r\in\mathcal R$, such as bandwidth, CPU cycles, GPU operations, or
memory capacity. Different commodities may impose different demands
on these resources. Consequently, the delay experienced on an edge
depends not only on the traffic rate, but also on how its resources
are divided among the commodities using it.

SPARQ distinguishes two execution regimes for representing this
interaction: Guaranteed-Resource (GR) and Shared-Resource (SR)
execution.

In the GR regime, illustrated in Fig.~\ref{fig:gr-model}, every
commodity receives a dedicated allocation of each required resource
and is served through commodity-specific queues. Its modeled
queueing delay is therefore determined by its own workload and
dedicated allocation, rather than by the instantaneous workloads of
other commodities. GR is representative of mechanisms such as
enforced container quotas, network slices, and dedicated virtual or
physical resources.

In the SR regime, illustrated in Fig.~\ref{fig:sr-model},
commodities using an augmented edge access a common capacity pool
for each resource. Capacity not used by one commodity remains
available to the others, enabling statistical multiplexing and
improving resource utilization. However, their delays become coupled: increasing the workload of one commodity can increase the
waiting time experienced by the other commodities sharing the same
resource. For example, a surge in avatar-inference traffic may
increase the delay of an audio-processing stream using the same
accelerator.

When a commodity requires several resources, the rate supported by
resource $r$ is determined by the ratio between the allocated
resource rate and the amount of that resource required per request.
The effective request service rate is limited by the smallest of
these ratios, corresponding to the bottleneck operation shown in
Figs.~\ref{fig:gr-model} and~\ref{fig:sr-model}.

In the original SPARQ formulation, the augmented-edge sets
$\mathcal E^{a,\mathrm{GR}}$ and
$\mathcal E^{a,\mathrm{SR}}$ are specified before optimization.
Thus, the execution regime associated with an augmented edge is
fixed. MusiSPARQ preserves the service representation, fixed
embedding, multi-resource requirements, and queueing interpretations
of SPARQ, but removes this static restriction for switchable edges.
At each orchestration epoch, it determines which eligible
class-specific commodities should remain in SR and which should
receive temporary GR isolation. The following section introduces the
subscription classes and formalizes this adaptive decision.

\begin{figure}[t]
    \centering

    \begin{subfigure}[t]{\linewidth}
        \centering
        \includegraphics[width=\linewidth]{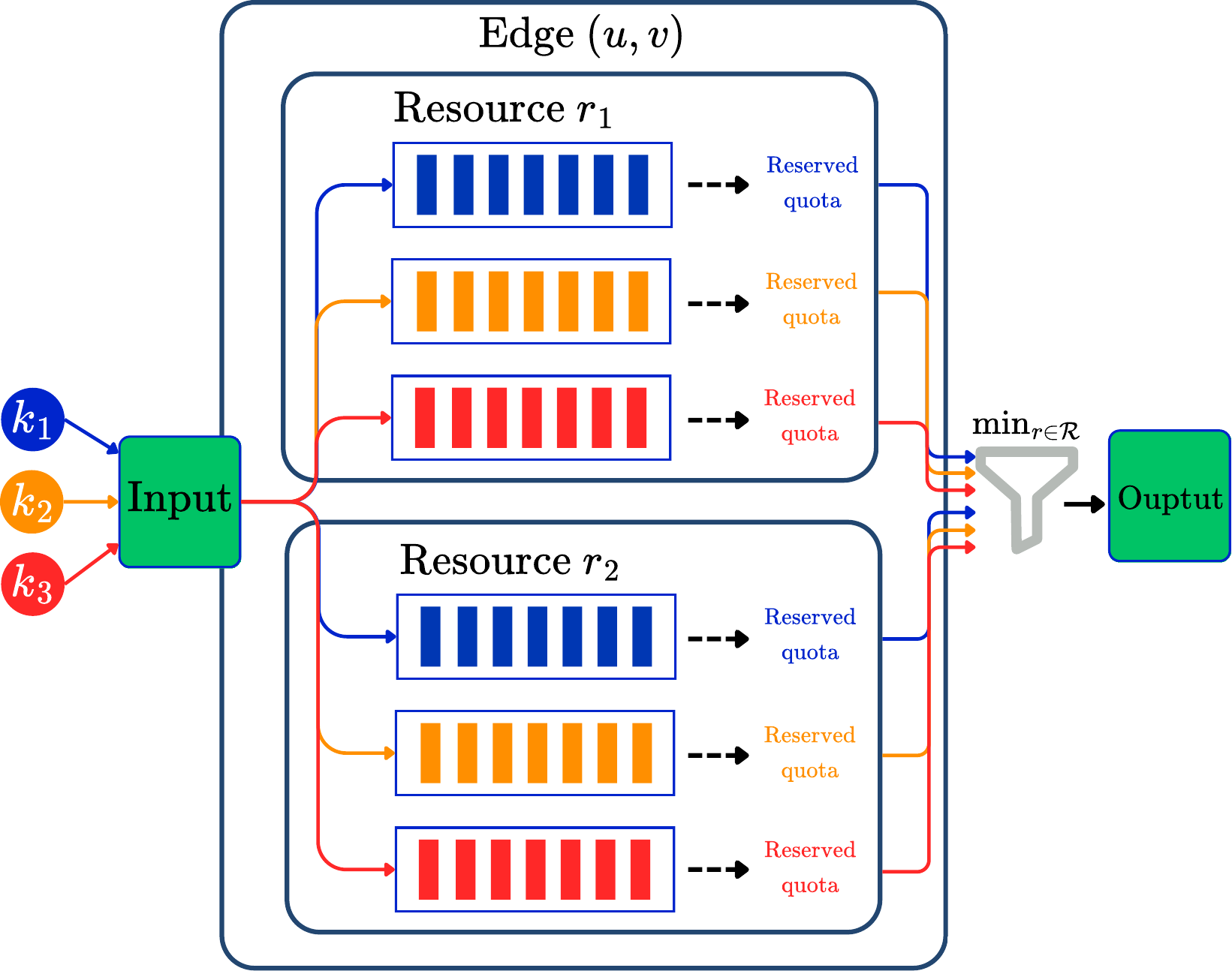}
        \caption{\footnotesize  Guaranteed-resource (GR) execution. Each commodity is
assigned independent queues and dedicated capacity for its required
resources, preventing modeled queueing interference from other
commodities.}
        \label{fig:gr-model}
    \end{subfigure}

    \vspace{0.4cm}

    \begin{subfigure}[t]{\linewidth}
        \centering
        \includegraphics[width=\linewidth]{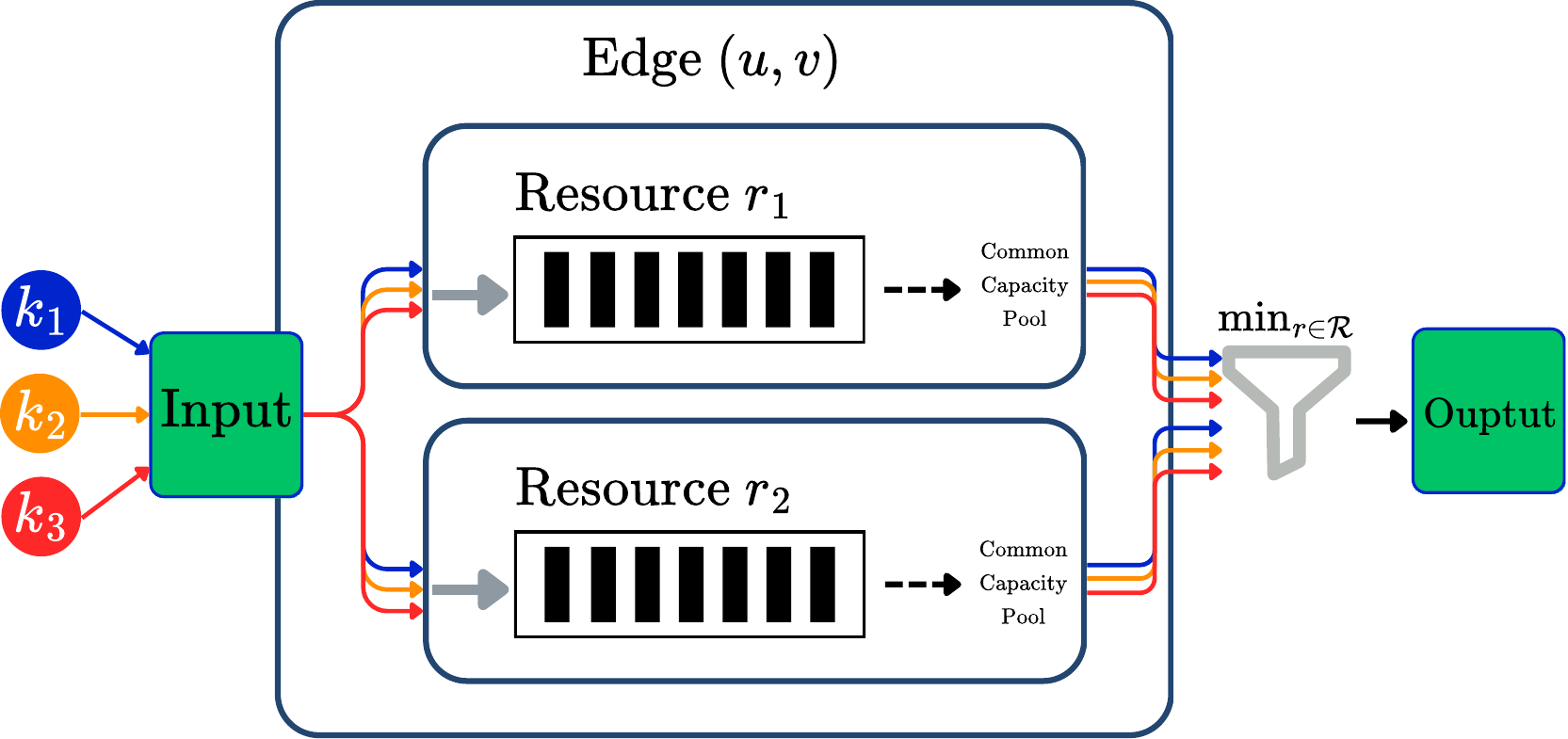}
        \caption{Shared-resource (SR) execution. Commodities share a
queueing system and capacity pool for each required resource, so
their delays depend on the aggregate workload.}
        \label{fig:sr-model}
    \end{subfigure}

    \caption{\footnotesize Execution regimes considered by MusiSPARQ. In GR
    execution, selected streams receive dedicated capacity and
    independent queueing service. In SR execution, streams share the available capacity and experience workload-dependent contention.
    MusiSPARQ dynamically assigns eligible component--class streams between these logical regimes while both draw from the fixed
    capacity of the same physical resource.}
    \label{fig:gr-sr-models}
\end{figure}

\section{MusiSPARQ System Model}
\label{sec:musisparq_system_model}

The preceding section described the service representation and the two execution regimes inherited from SPARQ. We now introduce the subscription-aware runtime decision made by MusiSPARQ. At the beginning of every orchestration epoch $t$, the controller observes the current request rates and determines which eligible commodities should remain in SR and which should temporarily receive GR isolation. Placement and routing remain fixed throughout this process.

\subsection{Subscription Classes and Runtime Traffic}

We consider the set of subscription classes $
    \mathcal{C}=\{1,\ldots,C\}
$. Examples include Premium, Standard, and Basic users. Each class $c \in \mathcal{C}$
is characterized by a preferred end-to-end latency target $L_c$, a
maximum admissible latency $L_c^{\max}$, a violation weight $w_c$,
and an isolation-eligibility parameter    $g_c\in\{0,1\}$ , with $g_c=1$ if class-$c$ traffic may receive isolated capacity.
 The preferred target represents the desired service quality, whereas
$L_c^{\max}$ provides a hard protection limit. The weight $w_c$
allows violations of latency-sensitive classes to receive greater importance in the orchestration objective.

Subscription-aware isolation requires distinguishing traffic from different classes even when it traverses the same deployed service functions. For example, Premium and Basic requests may both use the same audio-enhancement function, but MusiSPARQ may need to isolate only the Premium traffic during congestion. Representing both classes through one aggregate commodity would provide only one execution-mode decision and would therefore force them to remain together in either SR or GR.

Accordingly, in the MusiSPARQ runtime model, each index
$k\in\mathcal K$ identifies a class-specific instance of a deployed commodity. We use $c(k)\in\mathcal C$ to denote its subscription class. Thus, Premium audio and Basic audio are represented by different commodity indices, even though they may use the same service function and inherit the same fixed route $\bar{\mathbf f}$. This distinction is logical: it separates their traffic measurements, latency objectives, and isolation decisions without duplicating the physical function or recomputing its placement.

Let $\Lambda^c(t)$ denote the request rate generated by class $c$
during epoch $t$. For example, this rate may be estimated as
$\Lambda^c(t)=N_c(t)\eta_c$, where $N_c(t)$ is the number of active
class-$c$ users and $\eta_c$ is their average per-user request rate.

Because the SPARQ embedding is fixed, the request rate of commodity
$k$ on augmented edge $(u,v)$ is:
\begin{equation}
\lambda_{uv}^{k}(t)
=
\bar f_{uv}^{k}\Lambda^{c(k)}(t),
\label{eq:commodity-arrival}
\end{equation}

Therefore, an audience surge changes the traffic offered to the edges without changing the service placement or routes. 

Each augmented edge may provide several resource types
$r\in\mathcal R$. The parameter $R_{uv}^{k,r}$ denotes the amount of resource $r$ required to serve one request of commodity $k$ on edge $(u,v)$, while $M_{uv}^{r}$ denotes the available capacity of that resource. 

For convenience, we define
\begin{align}
\mathcal K_{uv}
&=\{k\in\mathcal K:\bar f_{uv}^{k}=1\},\\
\mathcal R_{uv}^{k}
&=\{r\in\mathcal R:R_{uv}^{k,r}>0\},
\end{align}
namely, the commodities traversing edge $(u,v)$ and the resources
required by commodity $k$ on that edge.

\subsection{Adaptive Execution Mode and Capacity Partition}

The central decision of MusiSPARQ is
\begin{equation}
z_{uv}^{k}(t)=
\begin{cases}
0, & \text{commodity $k$ uses SR on edge $(u,v)$},\\
1, & \text{commodity $k$ uses GR on edge $(u,v)$}.
\end{cases}
\label{eq:adaptive-mode}
\end{equation}

For example, Premium audio and Basic media traffic may initially share an edge GPU. During an audience surge, MusiSPARQ may isolate the Premium audio commodity while leaving Basic media traffic in the shared pool. When the surge ends, the audio commodity can return to SR, releasing its dedicated reservation.

The complete isolation configuration at epoch $t$ is
\begin{equation}
\mathbf z(t)=
\left\{
z_{uv}^{k}(t):
(u,v)\in\mathcal E^a,\;
k\in\mathcal K_{uv}
\right\}.
\end{equation}
The previous configuration $\mathbf z(t-1)$ is assumed known when the epoch-$t$ decision is computed.

Not every commodity can be isolated on every edge. The parameter $h_{uv}^{k}\in\{0,1\}$ indicates whether edge $(u,v)$ supports an independent queue and enforceable resource reservation for commodity $k$. Hence,
\begin{equation}
z_{uv}^{k}(t)
\leq
\bar f_{uv}^{k}h_{uv}^{k}g_{c(k)}.
\label{eq:isolation-eligibility}
\end{equation}
This condition combines deployment, platform, and subscription
eligibility.

For every eligible commodity, let
$\bar\mu_{uv}^{k,r,\mathrm{GR}}(t)$ denote the capacity of resource $r$ that must be reserved when the commodity is isolated. These requirements are runtime inputs obtained through profiling or a lower-level capacity-sizing mechanism. They may change across epochs as the offered traffic changes, but they are not independent optimization variables in MusiSPARQ. A valid GR requirement must provide a stability margin:
\begin{equation}
\lambda_{uv}^{k}(t)R_{uv}^{k,r}
\leq
\bar\rho^{\mathrm{GR}}
\bar\mu_{uv}^{k,r,\mathrm{GR}}(t),
\qquad
0<\bar\rho^{\mathrm{GR}}<1.
\label{eq:gr-sizing}
\end{equation}

Given an isolation configuration $\mathbf z(t)$, the capacity
remaining for SR is
\begin{equation}
\mu_{uv}^{r,\mathrm{SR}}(\mathbf z,t)
=
M_{uv}^{r}
-
\sum_{k\in\mathcal K_{uv}}
z_{uv}^{k}(t)
\bar\mu_{uv}^{k,r,\mathrm{GR}}(t).
\label{eq:residual-sr-capacity}
\end{equation}

This equation captures the physical coupling between the two
regimes. Isolating a commodity removes its workload from SR, but its
dedicated reservation simultaneously reduces the capacity available
to the commodities that remain shared.

The shared workload offered to resource $r$ is
\begin{equation}
A_{uv}^{r,\mathrm{SR}}(\mathbf z,t)
=
\sum_{k\in\mathcal K_{uv}}
(1-z_{uv}^{k}(t))
\lambda_{uv}^{k}(t)R_{uv}^{k,r}.
\label{eq:shared-workload}
\end{equation}
To maintain a stable operating margin, it must satisfy
\begin{equation}
A_{uv}^{r,\mathrm{SR}}(\mathbf z,t)
\leq
\bar\rho^{\mathrm{SR}}
\mu_{uv}^{r,\mathrm{SR}}(\mathbf z,t),
\qquad
0<\bar\rho^{\mathrm{SR}}<1.
\label{eq:sr-stability}
\end{equation}

The bottleneck service rates depicted in
Fig.~\ref{fig:gr-sr-models} are consequently
\begin{align}
\nu_{uv}^{k,\mathrm{SR}}(\mathbf z,t)
&=
\min_{r\in\mathcal R_{uv}^{k}}
\frac{\mu_{uv}^{r,\mathrm{SR}}(\mathbf z,t)}
     {R_{uv}^{k,r}},
\label{eq:sr-service-rate}\\
\nu_{uv}^{k,\mathrm{GR}}(t)
&=
\min_{r\in\mathcal R_{uv}^{k}}
\frac{\bar\mu_{uv}^{k,r,\mathrm{GR}}(t)}
     {R_{uv}^{k,r}}.
\label{eq:gr-service-rate}
\end{align}
The minimum appears because a request cannot be completed faster
than its slowest required resource.
Finally, changing the execution regime may require queue reassignment, container reconfiguration, or enforcement of a new resource reservation. We represent such changes through
\begin{equation}
s_{uv}^{k}(t)
=
\left|z_{uv}^{k}(t)-z_{uv}^{k}(t-1)\right|.
\label{eq:switching-indicator}
\end{equation}

\subsection{Queueing and End-to-End Delay}

We inherit the queueing assumptions and final delay expressions from
SPARQ~\cite{sparq}. Request arrivals are modeled as Poisson processes and the service discipline is FCFS. MusiSPARQ does not introduce a new queue model; it changes which inherited model applies to each commodity at each epoch.
When commodity $k$ uses GR, each required resource is represented
by an independent M/M/1 queue. Its mean sojourn time with respect
to resource $r$ is
\begin{equation}
d_{uv}^{k,r,\mathrm{GR}}(t)
=
\frac{R_{uv}^{k,r}}
{
\bar\mu_{uv}^{k,r,\mathrm{GR}}(t)
-
\lambda_{uv}^{k}(t)R_{uv}^{k,r}
}.
\label{eq:gr-delay}
\end{equation}
This delay depends only on the commodity's arrival rate and dedicated
allocation, and not on the traffic remaining in SR.

When commodity $k$ remains shared, the commodities using resource
$r$ form an M/G/1 queue. Their heterogeneous service requirements
produce the mean sojourn time
\begin{equation}
\small
\begin{split}
d_{uv}^{k,r,\mathrm{SR}}(\mathbf z,t)
={}&
\frac{
\displaystyle
\sum_{j\in\mathcal K_{uv}}
(1-z_{uv}^{j})
\lambda_{uv}^{j}(t)
\left(R_{uv}^{j,r}\right)^2
}{
\mu_{uv}^{r,\mathrm{SR}}(\mathbf z,t)
\left[
\mu_{uv}^{r,\mathrm{SR}}(\mathbf z,t)
-
A_{uv}^{r,\mathrm{SR}}(\mathbf z,t)
\right]
}
\\
&+
\frac{R_{uv}^{k,r}}
{\mu_{uv}^{r,\mathrm{SR}}(\mathbf z,t)}.
\end{split}
\label{eq:sr-delay}
\end{equation}

The first term represents the waiting time generated collectively by
the commodities remaining in SR. The second is the expected service
time of commodity $k$. Therefore, unlike GR, the SR delay of an
audio commodity may increase because avatar or media traffic becomes
more demanding.

Following the bottleneck representation used in SPARQ, the modeled
edge delay under each regime is
\begin{align}
d_{uv}^{k,\mathrm{SR}}(\mathbf z,t)
&=
\max_{r\in\mathcal R_{uv}^{k}}
d_{uv}^{k,r,\mathrm{SR}}(\mathbf z,t),\\
d_{uv}^{k,\mathrm{GR}}(t)
&=
\max_{r\in\mathcal R_{uv}^{k}}
d_{uv}^{k,r,\mathrm{GR}}(t).
\end{align}
These expressions represent the maximum of the per-resource mean
delays, following the tractable bottleneck approximation adopted by
SPARQ~\cite{sparq}.

The delay selected by the adaptive decision is
\begin{equation}
d_{uv}^{k}(\mathbf z,t)
=
(1-z_{uv}^{k})d_{uv}^{k,\mathrm{SR}}(\mathbf z,t)
+
z_{uv}^{k}d_{uv}^{k,\mathrm{GR}}(t).
\label{eq:selected-delay}
\end{equation}

The service graphs we consider are series--parallel: a class-$c$ request
traverses common sequential stages and, between them, parallel branches
that are synchronized at a merge point. Let $\mathcal P_c$ denote the
set of end-to-end processing paths of class $c$, each path $p$ being the
set of class-$c$ commodities traversed from source to delivery. Since a
merged output cannot be produced before its slowest input, the
end-to-end latency of class $c$ is determined by its slowest path:
\begin{equation}
D_c(\mathbf z,t)
=
B_c(t)
+
\max_{p\in\mathcal P_c}
\sum_{k\in p}
\sum_{(u,v)\in\mathcal E^a}
\bar f_{uv}^{k}\,d_{uv}^{k}(\mathbf z,t).
\label{eq:class-latency}
\end{equation}
For a purely sequential chain, $|\mathcal P_c|=1$ and the expression
reduces to the familiar sum of per-stage delays.

Here, $B_c(t)$ contains only latency contributions outside the
modeled augmented graph, such as synchronization or playout delay.
Communication and processing delays already represented by augmented
edges must not be included again in $B_c(t)$. The preferred-target violation is
\begin{equation}
\xi_c(\mathbf z,t)
=
[D_c(\mathbf z,t)-L_c]^+.
\label{eq:latency-violation}
\end{equation}
The hard constraint
\begin{equation}
D_c(\mathbf z,t)\leq L_c^{\max}
\label{eq:maximum-class-latency}
\end{equation}
prevents latency protection for one class from causing unacceptable
degradation to another.

\begin{table}[t]
\centering
\caption{Simulation parameters (P/S/B = Premium/Standard/Basic).}
\label{tab:sim_params}
\footnotesize
\setlength{\tabcolsep}{3.5pt}
\begin{tabular}{lll}
\toprule
Parameter & Symbol & Value \\
\midrule
Epochs $\times$ duration & --- & $300 \times 5$~s \\
Workload seeds & --- & $30$ \\
Preferred targets (P/S/B) & $L_c$ & $40/60/80$~ms \\
Maximum latencies (P/S/B) & $L_c^{\max}$ & $50/100/150$~ms \\
Violation weights (P/S/B) & $w_c$ & $4/2/1$ \\
GR eligibility (P/S/B) & $g_c$ & $1/1/0$ \\
Reference mix (P/S/B) & --- & $20/30/50\%$ \\
External delay (all classes) & $B_c$ & $20$~ms \\
Utilization limits & $\bar\rho^{\mathrm{SR}}=\bar\rho^{\mathrm{GR}}$ & $0.90$ \\
Reservation-cost coefficient & $\kappa$ & $0.5$ \\
Objective weights (default) & $\alpha,\ \beta$ & $200,\ 0.05$ \\
Base per-user rate & $\eta$ & $0.8$~req/s \\
Saturation audience & $N_{\mathrm{sat}}$ & $240$ users \\
\bottomrule
\end{tabular}
\end{table}

\subsection{Per-Epoch Isolation Decision}

At each epoch, MusiSPARQ balances three effects: the opportunity cost
of reserving capacity, subscription-specific latency violations, and
the overhead of changing execution regime. The isolation cost is 
\begin{equation}
C_{\mathrm{iso}}(\mathbf z,t)
=
\sum_{(u,v)\in\mathcal E^a}
\sum_{k\in\mathcal K_{uv}}
\sum_{r\in\mathcal R_{uv}^{k}}
\kappa_{uv}^{r}
z_{uv}^{k}(t)
\bar\mu_{uv}^{k,r,\mathrm{GR}}(t),
\label{eq:isolation-cost}
\end{equation}
where $\kappa_{uv}^{r}$ measures the opportunity or operational cost
of reserving resource $r$ instead of leaving it shared. The switching cost is 
\begin{equation}
C_{\mathrm{sw}}(\mathbf z,t)
=
\sum_{(u,v)\in\mathcal E^a}
\sum_{k\in\mathcal K_{uv}}
s_{uv}^{k}(t).
\label{eq:switching-cost}
\end{equation}

The per-epoch objective is
\begin{equation}
J(\mathbf z,t)
=
C_{\mathrm{iso}}(\mathbf z,t)
+
\alpha\sum_{c\in\mathcal C}w_c\xi_c(\mathbf z,t)
+
\beta C_{\mathrm{sw}}(\mathbf z,t),
\label{eq:per-epoch-objective}
\end{equation}
where $\alpha$ controls the importance of latency protection and
$\beta$ discourages unnecessary oscillations. MusiSPARQ solves 
\begin{equation}
\begin{aligned}
\min_{\mathbf z(t)}\quad
&J(\mathbf z,t)\\
\mathrm{s.t.}\quad
&z_{uv}^{k}(t)
 \leq\bar f_{uv}^{k}h_{uv}^{k}g_{c(k)}, &&(u,v),k,\\
&\sum_{k\in\mathcal K_{uv}}
 z_{uv}^{k}(t)\bar\mu_{uv}^{k,r,\mathrm{GR}}(t)
 \leq M_{uv}^{r},
 &&(u,v),r,\\
&A_{uv}^{r,\mathrm{SR}}(\mathbf z,t)
 \leq\bar\rho^{\mathrm{SR}}
 \mu_{uv}^{r,\mathrm{SR}}(\mathbf z,t),
 &&(u,v),r,\\
&z_{uv}^{k}(t)\lambda_{uv}^{k}(t)R_{uv}^{k,r}
 \leq
 \bar\rho^{\mathrm{GR}}z_{uv}^{k}(t)
 \bar\mu_{uv}^{k,r,\mathrm{GR}}(t),
 &&(u,v),k,r,\\
&D_c(\mathbf z,t)\leq L_c^{\max},
 &&c\in\mathcal C,\\
&z_{uv}^{k}(t)\in\{0,1\}.
\end{aligned}
\label{eq:musisparq-problem}
\end{equation}

All capacities and delays in this problem are induced by
$\mathbf z(t)$; they are not additional decisions. Consequently,
MusiSPARQ remains an adaptive mode-selection layer rather than a new
placement, routing, or continuous resource-allocation framework.

The controller evaluates all configurations over the eligible commodity--edge
pairs. For each candidate, it computes the induced SR capacity,
checks GR and SR stability, evaluates the class latencies, and
discards candidates violating a hard constraint. The feasible candidate with the smallest objective is applied during epoch $t$.

\section{Experimental Setup}
\label{sec:experimental_setup}

\paragraph*{Simulation environment}
We evaluate MusiSPARQ using an epoch-based simulator that implements the
model of Section~\ref{sec:musisparq_system_model}\footnote{The simulator code and workload traces are available at
\url{https://github.com/alialhousseini/musisparq}.}. Placement and routing are
fixed by design: all policies operate on the same embedded service
graph, so that performance differences are attributable solely to the
runtime SR/GR decision rather than to placement or routing effects. At
the beginning of each epoch, each policy observes the class-specific
request rates and selects its SR/GR configuration; the per-function
delays induced by this configuration are then computed from the
closed-form M/M/1 and M/G/1 expressions of
Section~\ref{sec:musisparq_system_model}. The reported latency therefore
corresponds to the modeled mean end-to-end delay in each epoch, rather
than to packet-level tail latency.

\paragraph*{Scenario and service model}
We consider a single edge site supporting audience interaction during a
live virtual concert~\cite{turchet2023musical}. Musicians provide the
concert content, while the simulated users are Premium, Standard, and
Basic audience members. Following representative MM
workflows~\cite{al2025mumenet}, the deployed service graph
(Fig.~\ref{fig:pipeline}) instantiates the commodity abstraction of
Section~\ref{sec:sparq_preliminaries}: audience sound and gesture data
traverse two parallel processing paths, an audio path (audience-sound
mixing followed by spatial-audio rendering) and a visual path
(avatar-motion inference followed by avatar synchronization), which
merge before personalized media delivery. All four functions execute on
a shared edge accelerator exposing two resource types, processing
throughput and memory bandwidth, whose capacities are normalized to one.
Per-request demands $R_{uv}^{k,r}$ are expressed as fractions of these
capacities; they are heterogeneous across functions and fixed across
experiments. This normalization decouples the results from any specific
hardware model and allows load to be read directly as resource
utilization. Capture, encoding, and delivery are modeled as additional non-switchable
single-resource stages traversed by all classes, with unit normalized
capacity; only the four GPU functions are eligible for isolation
($h_{uv}^{k}=1$). Access-network and playout delays are collected in the
constant external term $B_c$. Note that, under
Eq.~\eqref{eq:class-latency}, isolating a function reduces class latency
only if it lies on the slowest path; the per-epoch optimization accounts
for this when selecting which commodities to protect.

\paragraph*{Subscription classes}
The three classes are differentiated by their preferred latency targets,
$L_{\mathrm P}=40$~ms, $L_{\mathrm S}=60$~ms, and $L_{\mathrm B}=80$~ms,
selected as service-level objectives for audience
interaction~\cite{shen2023toward, rottondi2016overview, boem2025issues,
turchet2023latency}, together with class-specific maximum admissible
latencies $L_c^{\max}$ and violation weights $w_c$ that assign greater
importance to higher tiers (Table~\ref{tab:sim_params}). Premium and Standard commodities are
eligible for GR isolation ($g_{\mathrm P}=g_{\mathrm S}=1$), whereas
Basic commodities always remain in SR ($g_{\mathrm B}=0$). The reference
audience composition is $20\%$ Premium, $30\%$ Standard, and $50\%$
Basic.

\begin{figure}
    \centering
    \includegraphics[width=0.9\linewidth]{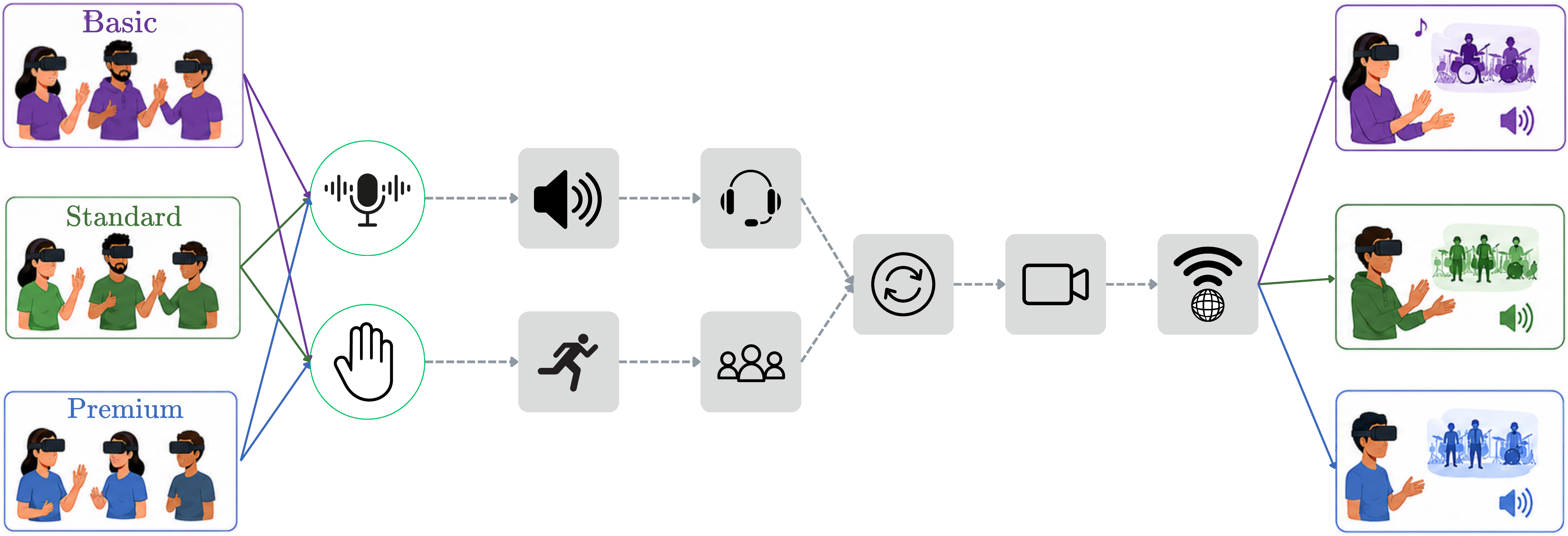}
    \caption{\footnotesize MM service graph used in the evaluation. Audience sound and gesture data traverse parallel audio (sound mixing, spatial-audio rendering) and visual (avatar-motion inference, avatar synchronization) paths that merge before personalized media delivery. All four processing functions share a common edge accelerator.}
    \label{fig:pipeline}
\end{figure}

\paragraph*{Workload and experiments}
Audience size is normalized by $N_{\mathrm{sat}}$, the size at which the
all-SR configuration reaches its safe operating limit
$\bar\rho^{\mathrm{SR}}$ under the reference mix; we vary the load as
$N=\gamma N_{\mathrm{sat}}$ with $\gamma\in[0.4,1]$. This makes the
results depend on the modeled capacity and per-user workload rather than
on an arbitrary user count. Three experiments are conducted. (i)~The
\emph{dynamic trace} contains 300 five-second epochs representing a
25-minute concert segment; as summarized in
Table~\ref{tab:workload_phases}, its phases capture audience admission,
regular performance, transient crowd-reaction bursts, a sustained
headline peak, and departure, where admission and departure modify the
active population while crowd reactions increase the per-user request
rate. (ii)~The \emph{audience-size sweep} varies $\gamma$ from $0.4$ to $1$
under the reference mix and, as a robustness check, additionally varies
the Premium share from $10\%$ to $70\%$ at $0.9N_{\mathrm{sat}}$.
(iii)~The \emph{sensitivity study} varies the violation weight $\alpha$
over a logarithmic grid around the default configuration ($\alpha=200$,
$\beta=0.05$) to characterize the latency--cost operating curve.

\paragraph*{Baseline approaches}
We compare MusiSPARQ with three policies. \emph{All-SR} keeps all
commodities shared and represents maximum statistical multiplexing.
\emph{Static Premium} isolates every eligible Premium commodity
throughout the experiment and represents permanent subscription-based
reservation. \emph{Max-Feasible GR} selects, at each epoch, the feasible
configuration reserving the greatest GR capacity and represents
aggressive isolation. All policies use identical workload traces,
resource capacities, GR-sizing inputs, and feasibility constraints.

\paragraph*{Metrics and statistical protocol}
We report the mean end-to-end latency, preferred-target violation rate, and hard-limit violation rate of each class, together with the normalized GR capacity reserved by each policy. An epoch is infeasible for a policy when no
configuration available to that policy satisfies the stability and
hard-latency constraints; infeasible epochs are counted as a separate
metric and excluded from the latency statistics of the affected policy.
Each experiment is repeated over 30 independent workload seeds, with
identical seeds across policies.

\begin{table}[t]
\centering
\caption{\footnotesize  Dynamic virtual-concert workload. Load is expressed as
all-SR resource utilization before isolation.}
\label{tab:workload_phases}
\begin{tabular}{ccl}
\hline
Epochs & Load & Event phase \\
\hline
1--40    & $0.35$--$0.55$ & Audience admission \\
41--90   & $0.55$          & Opening performance \\
91--120  & $0.80$          & Crowd-reaction burst \\
121--180 & $0.65$          & Main performance \\
181--240 & $0.90$          & Headline/encore period \\
241--300 & $0.60$--$0.30$ & Recovery and departure \\
\hline
\end{tabular}
\end{table}

\section{Results and Discussion}
\label{sec:results}

We organize the discussion around three questions. First, does adaptive
isolation protect Premium latency during a realistic dynamic event, and
at what resource cost (Figs.~\ref{fig:dynamic_latency}
and~\ref{fig:load_reservation}). Second, how does the violation weight
$\alpha$ shape the latency--cost operating point, and where does the
default configuration sit (Fig.~\ref{fig:cost_protection}). Third, does
the observed behavior persist as the system approaches saturation
(Fig.~\ref{fig:audience_scalability})? Each experiment ends with a short
takeaway summarizing the lesson it supports.

\subsection{Dynamic Latency Protection}

Fig.~\ref{fig:dynamic_latency} shows the seed-averaged end-to-end
latency of each class under the four policies during
the two congestion-prone portions of the virtual concert, with the
preferred targets $L_c$ marked by dashed lines. During the unshaded
intervals, all policies keep all three classes below their targets. The
behavior diverges inside the shaded crowd-reaction and headline phases:
the All-SR Premium latency rises above the $40$~ms target and remains
there for most of both bursts, whereas Static Premium, Max-Feasible GR,
and MusiSPARQ keep the Premium curve below the target throughout.
Aggregating over the trace, All-SR exhibits a Premium target-violation
rate of $16.19\%$, while MusiSPARQ achieves a rate of $0.29\%$, i.e.,
$98.22\%$ lower. Beyond the mean curves shown in the figure, the
per-epoch statistics also record hard-limit violations: All-SR exceeds
the Premium hard limit $L_{\mathrm P}^{\max}$ in $0.63\%$ of epochs,
whereas MusiSPARQ produces none.

The middle and bottom panels of Fig.~\ref{fig:dynamic_latency} reveal
the side effect of permanent isolation. By removing capacity from the
shared pool even when protection is unnecessary, Static Premium pushes
Standard latency above its target during the bursts, yielding
target-violation rates of $9.13\%$ for Standard and $1.54\%$ for Basic,
together with hard-limit violations in $0.51\%$ and $0.17\%$ of epochs,
respectively. MusiSPARQ instead keeps all three classes within their
preferred targets: its Premium protection is not paid for by the lower
tiers.

\begin{figure}
    \centering
    \includegraphics[width=0.9\linewidth]{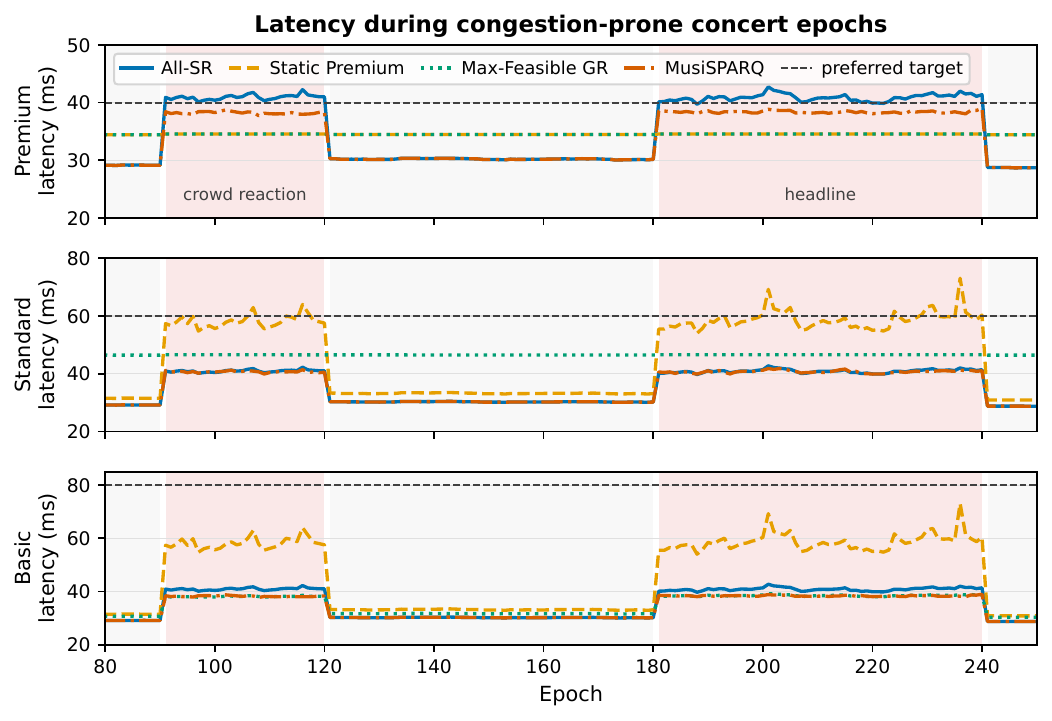}
\caption{\footnotesize  Dynamic latency during congestion-prone concert epochs. Dashed horizontal lines indicate the
preferred latency target of each class. Shaded regions mark the
crowd-reaction and headline phases, where transient interaction bursts
increase shared-resource contention.}
\label{fig:dynamic_latency}
\end{figure}

\paragraph*{Takeaway-1} Under transient congestion, both extremes of
the sharing--isolation spectrum fail in complementary ways: full sharing
sacrifices the latency-critical class, while permanent isolation
sacrifices the remaining ones. Protecting a class only \emph{when} it is
threatened avoids both failure modes.

To understand the resource cost at which this protection is obtained,
Fig.~\ref{fig:load_reservation} pairs the workload pressure with the
reservation decisions over the full trace. The upper panel reports the
all-SR load before isolation, i.e., the maximum accelerator utilization
that would occur if all traffic remained shared; the crowd-reaction and
headline phases reach average loads of $84.71\%$ and $84.75\%$, whereas
the opening and main-performance phases remain at $58.21\%$ and
$64.01\%$. The lower panel reports the GR capacity reserved by each
policy. Static Premium and Max-Feasible GR hold their reservations for
the entire event, averaging $20.96\%$ and $34.31\%$ of capacity. The
MusiSPARQ curve instead follows the load: it is zero outside the shaded
phases, rises to an average of $2.44\%$ during the crowd-reaction burst
and $2.31\%$ during the headline period, and returns to zero when each
burst ends. Its trace-wide average is $0.71\%$, corresponding to
$96.64\%$ and $97.94\%$ less reserved capacity than Static Premium and
Max-Feasible GR, respectively.

\paragraph*{Takeaway-2} Matching reservation to congestion, rather than
to subscription class alone, is sufficient: brief, small, and reversible
reservations during load bursts deliver the protection for which static
policies pay continuously.

\begin{figure}
    \centering
    \includegraphics[width=0.9\linewidth]{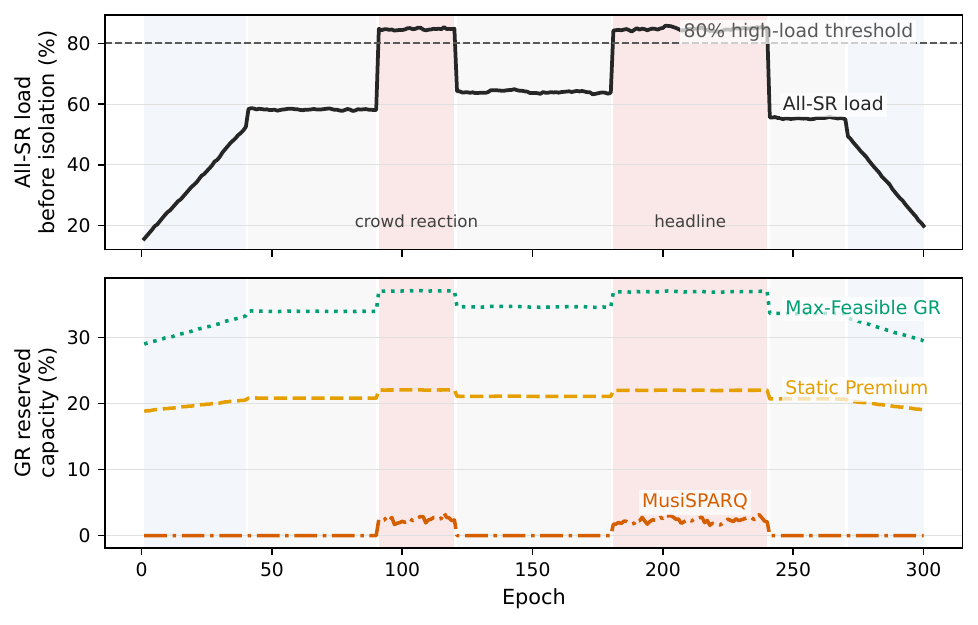}
\caption{\footnotesize  Workload pressure and GR reservation over the dynamic
virtual-concert trace. The upper panel shows the all-SR load before
isolation, computed as the maximum edge-accelerator utilization that
would occur if all traffic remained in SR. The lower panel reports the
normalized GR capacity reserved by each policy. Shaded regions denote
the crowd-reaction and headline phases.}
\label{fig:load_reservation}
\end{figure}

\subsection{Latency--Cost Operating Point}

The dynamic-trace results were obtained with the default configuration
($\alpha=200$, $\beta=0.05$); we now examine how the violation weight
$\alpha$ determines this operating point and how an operator can tune
it. In the objective, $\alpha$ prices unresolved preferred-target
violations: small values favor low reservation cost, while large values
make MusiSPARQ increasingly willing to reserve GR capacity to suppress
them.

Fig.~\ref{fig:cost_protection}(a) sweeps $\alpha$ over a logarithmic
grid. At $\alpha=1$, MusiSPARQ reserves almost nothing ($0.03\%$ of
capacity) and its Premium violation rate, $16.19\%$, matches that of
All-SR: when violations are priced too low, the optimal decision is
simply not to isolate. As $\alpha$ grows, the violation rate falls
rapidly at nearly constant reservation: to $5.11\%$ at $\alpha=10$ (with
$0.54\%$ reserved) and to $0.56\%$ at $\alpha=100$ (with $0.70\%$
reserved), a roughly $29$-fold reduction relative to $\alpha=1$ obtained
with under one percent of the capacity. Beyond $\alpha\approx100$ the
curve flattens: further increases buy only marginal latency gains, which
identifies a natural operating region. The default $\alpha=200$ lies in
this region, between the $\alpha=177.8$ and $\alpha=316.2$ grid points,
and its $0.29\%$ violation rate at $0.71\%$ reservation is consistent
with the surrounding samples.

We now place this default operating point among the fixed baselines,
shown in Fig.~\ref{fig:cost_protection}(b). All-SR occupies the
zero-reservation endpoint with a $16.19\%$ Premium violation rate;
Static Premium and Max-Feasible GR occupy the protection endpoints,
eliminating Premium violations at $20.96\%$ and $34.31\%$ average
reservation. MusiSPARQ sits between them, at $0.29\%$ violation and
$0.71\%$ reservation.

\paragraph*{Takeaway-3} MusiSPARQ reaches essentially the same Premium
protection as the isolation-heavy baselines ($0.29\%$ violation) while
reserving $0.71\%$ of capacity instead of $20.96\%$ or $34.31\%$ --
protection at roughly $30$--$50\times$ less reserved capacity. This
contrast is the quantitative form of the missing middle ground between
purely shared and purely isolated operation, and $\alpha$ gives the
operator direct control over where on this curve to sit.
\begin{figure}
    \centering
    \includegraphics[width=0.9\linewidth]{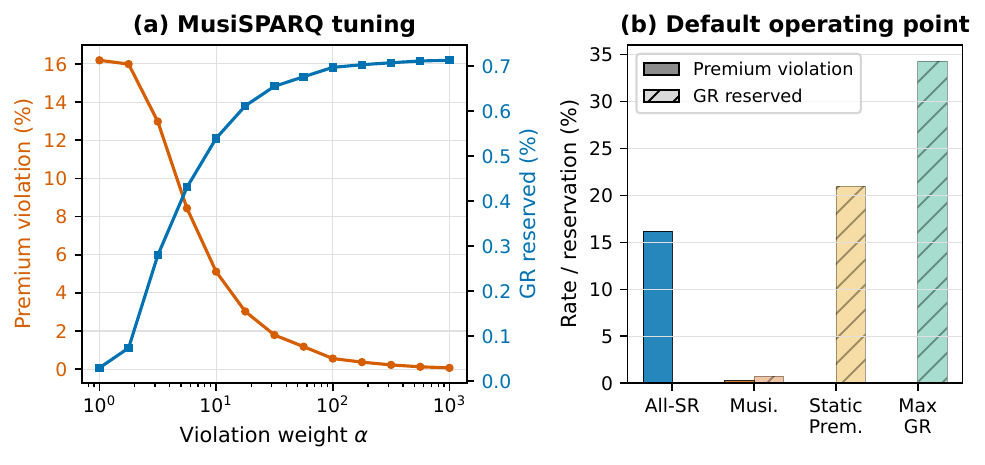}
    \caption{\footnotesize  Latency--cost operating behavior of MusiSPARQ. Panel~(a) shows how the violation weight $\alpha$ controls the tradeoff between Premium preferred-target violations and GR reservation. Panel~(b) compares the selected default MusiSPARQ operating point with the fixed baselines in terms of Premium violation rate and reserved GR capacity.}
    \label{fig:cost_protection}
\end{figure}

\begin{figure}
    \centering
    \includegraphics[width=0.9\linewidth]{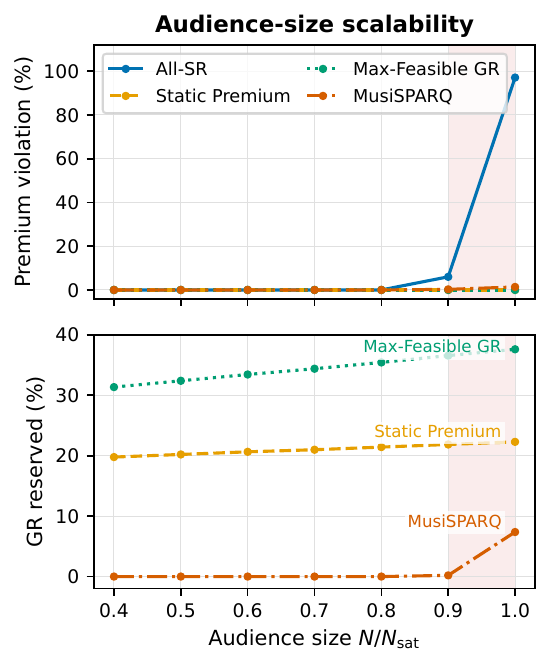}
    \caption{\footnotesize  Audience-size scalability under the reference subscription mix. The audience size is varied from $0.4N_{\mathrm{sat}}$ to $N_{\mathrm{sat}}$. The upper panel reports the Premium preferred-target violation rate, while the lower panel reports the normalized GR capacity reserved by each policy. The shaded region marks operation close to saturation.}
    \label{fig:audience_scalability}
\end{figure}

\subsection{Audience-Size Scalability}

The experiments so far operate around the reference load; we finally ask
whether adaptive isolation remains effective as the system is pushed
toward saturation. Fig.~\ref{fig:audience_scalability} varies the
audience size from $0.4N_{\mathrm{sat}}$ to $N_{\mathrm{sat}}$, with the
upper panel reporting the Premium violation rate and the lower panel the
reserved GR capacity.

For moderate audiences, from $0.4N_{\mathrm{sat}}$ to
$0.8N_{\mathrm{sat}}$, the upper panel shows no Premium violations for
either All-SR or MusiSPARQ, and the lower panel shows that MusiSPARQ
reserves no capacity in this whole region -- its behavior coincides with
All-SR because sharing alone satisfies every target. Static Premium and
Max-Feasible GR nevertheless hold $19.78$--$21.43\%$ and
$31.36$--$35.44\%$ of capacity: reservation spent without any latency
benefit. The regimes separate near saturation. At $0.9N_{\mathrm{sat}}$,
All-SR reaches a $6.00\%$ Premium violation rate, which MusiSPARQ limits
to $0.27\%$ using only $0.20\%$ reservation. At $N_{\mathrm{sat}}$,
All-SR collapses -- $97.07\%$ Premium violations and $45.53\%$
infeasible epochs -- while MusiSPARQ contains violations to $1.37\%$
with $7.37\%$ reservation and no infeasible epochs; Static Premium and
Max-Feasible GR also avoid violations but at $22.29\%$ and $37.61\%$
reservation. As a complementary robustness check, we fixed the audience
at $0.9N_{\mathrm{sat}}$ and swept the Premium share from $10\%$ to
$70\%$: MusiSPARQ keeps the Premium violation rate below $1\%$
($0.27$--$0.53\%$) with at most $0.38\%$ reservation, whereas the
reservation of the static policies grows proportionally with the Premium
share (results not shown).

\paragraph*{Takeaway-4} Adaptive isolation degrades gracefully where
fixed policies fail abruptly: it stays indistinguishable from full
sharing while sharing suffices, and injects just enough reservation near
saturation to prevent the collapse of shared execution -- at a fraction
of the capacity that static protection requires.

\section{Conclusion}
\label{sec:conclusion}

This paper introduced MusiSPARQ, a subscription-aware control framework that promotes the shared-versus-guaranteed execution regime of SPARQ from a fixed infrastructure attribute to a per-epoch runtime decision. By representing each switchable resource through coupled SR and GR pools drawing from a common capacity budget, MusiSPARQ selectively isolates class-specific service flows when contention threatens their latency objectives and returns the reserved capacity to shared execution once congestion subsides. The per-epoch decision jointly accounts for latency violations, isolation cost, switching overhead, and per-class maximum-latency limits. In a dynamic virtual-concert scenario, MusiSPARQ reduced the Premium latency-target violation rate from 16.19\% to 0.29\% while reserving on average only 0.71\% of GR capacity and, unlike static isolation, keeping Standard and Basic traffic within their latency targets. Future work includes scalable selection heuristics for larger deployments, predictive isolation driven by workload forecasting, and joint adaptation of the execution regime with placement and routing decisions.

\bibliographystyle{IEEEtran}
\bibliography{references}

\end{document}